\documentclass[aps,prb,twocolumn,showpacs,floats]{revtex4}
\usepackage{graphicx}%
\usepackage{epsfig}
\usepackage{amsmath}%
\usepackage{amssymb}%

\def\beq{\begin{equation}}
\def\eeq{\end{equation}}
\def\e{\epsilon}

\def\beq{\begin{equation}}
\def\eeq{\end{equation}}
\def\beqa{\begin{eqnarray}}
\def\eeqa{\end{eqnarray}}
\def\half{\frac{1}{2} }

\def\e{\epsilon}
\def\half{{\ss 1\over 2}}

\def\si{\sigma}
\def\del{\delta }
\def\D{\Delta}

\def\ss{\scriptstyle}

\def\half{{\ss 1\over 2}}

\def\si{\sigma}
\def\del{\delta}

\def\ss{\scriptstyle}

\def\prl{{\sl Phys. Rev. Lett.}\ }
\def\Etip{E_\mathrm{tip}}
\def\leff{l_\mathrm{eff}}
\begin{document}
\bibliographystyle{prsty}
\title{\Large\bf  }\title{\Large\bf Local Current Distribution and "Hot Spots" in the Integer Quantum Hall Regime }
\author{Yonatan Dubi$^1$, Yigal Meir$^{1,2}$ and Yshai Avishai$^{1,2,3}$}
\affiliation{
$^{1}$ Physics Department, Ben-Gurion University, Beer Sheva 84105, Israel\\
$^{2}$ The Ilse Katz Center for Meso- and Nano-scale Science and
Technology, Ben-Gurion University, Beer Sheva 84105, Israel \\
$3$ Department of Applied Physics, University of Tokyo, Hongo Bunkyo-ku, Tokyo 113, Japan}
\date{\today}
\begin{abstract}
In a recent experiment, the local current distribution of a
two-dimensional electron gas in the quantum Hall regime was probed
by measuring the variation of  the conductance due to local
gating. The main experimental finding was the existence of "hot
spots", i.e. regions with high degree of sensitivity to local
gating, whose density increases as one approaches the quantum Hall
transition. However, the direct connection between these "hot
spots" and regions of high current flow is not clear. Here, based
on a recent model for the quantum Hall transition consisting of a
mixture of perfect and quantum links, the relation between the
"hot spots" and the current distribution in the sample has been
investigated. The model reproduces the observed dependence of the number
and sizes of "hot spots" on the filling factor. It is further
demonstrated that these "hot spots" are not located in regions
where most of the current flows, but rather, in places where the
currents flow both when injected from the left or from the right.
A quantitative measure, the harmonic mean of these currents is
introduced and correlates very well with the "hot spots"
positions.
\end{abstract}

 \maketitle
\section{Introduction}
The quantum Hall (QH) effect remains a major focus of
interest,\cite{QHEreview} despite the long time that has passed
since its discovery. Apparently this is due to the ongoing
technological progress employing new experimental probes and
yielding new and sometimes surprising results. A particular issue,
that has been under debate for quite some time, is related to the
exact trajectories at which the current flows. Some theories suggest
that the current flows mainly via edge states along the sample
edges,\cite{edge} while others, based on the idea of a
localization-delocalization transition at the centers of Landau
levels, predict a distribution of currents extending throughout
the bulk.\cite{bulk} However, due to the robustness of the
conductance quantization, the local properties are inaccessible
via standard transport measurements. While some information on the
current flow can be derived from static probes, \cite{palevski}
numerous attempts have been made to address these questions using
various scanning imaging techniques,\cite{imaging} commonly based
on local probe of charge and electric potential. Yet, while
proving successful in describing localized electronic states,
these methods detect current only indirectly and cannot determine
unambiguously how the current is partitioned between the edge and
bulk channels.

Recently, a novel experimental approach has been applied to probe
 the local current distribution in a ballistic quantum point contact (QPC).\cite{Topinka}
An atomic force microscope (AFM) tip was placed on top of the 2DEG
in which a point contact was defined, causing a local depletion of
electron density beneath it. The underlying assumption in this
experiment is that this depletion strongly affects the conductance
through the QPC only if the current density under the tip is high.
On the other hand, the conductance should not be modified if there
is low current density under the tip. Thus, plotting the
conductance change as a function of tip position results in an
imaging of the electron current density. Indeed, the imaging
clearly showed the different modes of the electronic wave-function
being successively occupied as the conductance through the QPC
increases in quantized steps. Following this experiment a theoretical
model was devised \cite{metalidis} which mimics this experiment and yields similar
results for the distribution of current.

A similar experimental method has been used more recently to study the local
current distribution in a 2DEG in the QH regime.\cite{enslin} An
AFM tip was placed on top of the Hall bar and locally gated the sample,
 thus changing the local potential beneath it.  The
resistance was then measured as a function of tip position and
magnetic field. The main finding was that there are "hot spots" in
the sample, i.e. isolated regions at which the conductance is
extremely sensitive to the gating potential.  These domains were
interpreted as places where the current passes. They are mainly
observed in the transition between plateaus, and disappear almost
completely in the quantum Hall regime.

The fact that these measurement may not directly reflect the
total current distribution in the sample can be understood through
a simple example. Let us assume that there is a high barrier
separating the left side and the right side of the sample. Then a
current injected from either side (left or right) will be
localized on that side of the barrier. Nevertheless, the only
place where a change in the potential may induce a change in the
conductance of the system is at the barrier itself, where no
current actually flows. To put it in different terms, how can the
measurement of the conductance, a quantity that obeys specific
Onsager relations with respect to reversing directions of the
current and the magnetic field, yield information on the current
distribution which, to the best of our knowledge, do not obey
such symmetry relations?

Motivated by these experimentally relevant questions we employ in
this work a recently proposed model for the QH transition\cite{us}
that enables a theoretical modelling of the experiment and allows
for a detailed analysis of both the hot spots and of the spatial
distribution of currents in the sample. Our main finding, beyond a
good qualitative agreement with the experimental results, is that
the hot spots are located at points where current passes both when
it is injected from the left or from the right, and not
necessarily at points where the current density is high. The
number of such symmetry points is enhanced near the QH transition,
 but is small on either side of it, in agreement with
the experimental results. We propose an empirical relation
between the current distribution and the
position of the "hot spots", based on a harmonic average of
the current distributions when the current is injected from the left
and from the right.

\section{Model}
For the sake of completeness, let us briefly explain our model.
In strong magnetic fields, electrons with Fermi energy
$\epsilon_F$ perform small oscillations around equipotential
lines. When $\epsilon_F$ is small, their trajectories are trapped
inside potential valleys, with weak tunneling occurring between
adjacent valleys. We associate each such potential valley with a
site in a lattice. Nearest neighbor valleys (localized orbits) are
connected by links representing quantum tunneling between them. As
$\e_F$ increases and crosses the saddle point energy separating
two neighboring valleys, the two isolated trajectories coalesce,
the electron can freely move from one valley to its neighbor and
the link connecting them becomes perfect. The QH transition occurs
when an electron can traverse the sample along an equi-potential
trajectory (an edge-state), which in the model corresponds to
percolation of perfect links.

\begin{figure}[!h]
\centering
\includegraphics[width=8.4truecm]{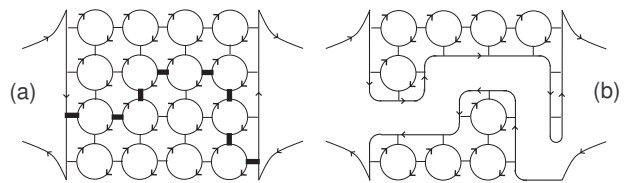}
\caption{Mapping of the problem onto a quantum percolation on a lattice: each
link carries two counter propagating edge modes (a). A non zero
transmission (thin lines) allows electrons to tunnel between
adjacent sites (potential valleys). When the transmission is unity
(bold lines in (a)), these two valleys merge, and an edge
state can freely propagate from
 one to another. A percolation of these perfect transmission links (b)
  correspond to an edge state propagating through
the system and a quantized conductance.
}
\label{Smatrixlattice2}
\end{figure}

Consequently, the QH problem maps onto a mixture of perfect and
quantum links on a lattice.  Each link carries a left going and a
right going channel. In accordance with the physics at strong
magnetic fields there is no scattering in the junctions (valleys)
and the edge state continues propagating uninterrupted according
to its chirality (see Fig.~\ref{Smatrixlattice2}(a)), while the
scattering occurs on the link (saddle point) itself. Each
scatterer is characterized by its scattering matrix $S_i$, namely
a transmission  probability $T_i$ and phases. The phases are taken
as random numbers from $0$ to $2 \pi$ and the transmission
amplitude of each link is determined by the height of the saddle
point barrier between the neighboring valleys, taken form a
uniform distribution $U [-V,V]$. The transmission is then
determined locally by the local barrier height $\e$ and the Fermi
energy $\e_F$ by \beq T(\e_F)=\exp\left[-\alpha (\e-\e_F)\right],
\label{TE} \eeq for $\e_F<\e$, where $\alpha$ is some constant,
and $T(\e_F)=1$ for $\e_F>\e$. The transmission through the
system, $T$, is then calculated using the scattering matrix
approach and the conductance of the system is determined by the
Landauer formula, $G=\frac{2 e^2}{h} T$.

\begin{figure}[!h]
\centering
\includegraphics[width=4truecm]{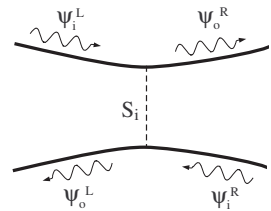}
\caption{Within the scattering model, each link is characterized by a scattering matrix $S_i$, connecting
between the incoming and outgoing wave functions from left to right (see text).}
\label{link}
\end{figure}

Within the scattering model, the current carried by the $i$-th
link is given by $J_i \propto | \psi_i^L |^2-| \psi_o^L |^2$, where
$\psi_{i (0)}^{L (R)}$ is the incoming (outgoing) wave function
from the left (right) of the link (see Fig~\ref{link}). Note that
due to the unitarity of the S-matrix, the current is a locally
left-right symmetric quantity, and is thus a property of the
entire link.

\begin{figure}[h!]
\centering
\includegraphics[width=7truecm]{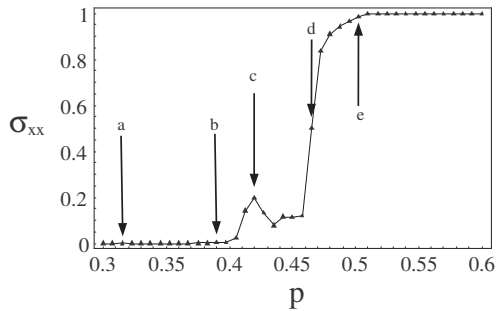}
\centering \caption{\footnotesize Conductance as a function of
probability $p$. The arrows point on the concentrations at which
the current distribution (Fig.~\ref{currents}) and the hot spots
(Fig.~\ref{HotLinks}) are plotted.}
\label{hotspots1} \end{figure}

As the Fermi energy increases, the conductance rises from zero to
unity at the percolation threshold (Figs.~\ref{Smatrixlattice2}(b) and \ref{hotspots1}).
 In Ref.~\onlinecite{us} it was demonstrated that the phase transition described by the above model
exhibits a critical exponent $\nu\simeq 2.4$, in agreement with
numerical simulations for other  models describing the QH
transition. In this paper we simulate the experiment of
Ref.~\onlinecite{enslin} using the above model.

\section{Results}

\subsection{Current Distribution}

In Fig.~\ref{hotspots1} the conductance (in units of
$2e^2/h$) of a $20 \times 20$ size system \cite{square} is plotted as a function of
probability of a link to be perfect which corresponds to Fermi
energy, $p=\half \left( \frac{\epsilon_F}{V}+1 \right)$. As can be seen, our
model reproduces the conductance fluctuations close to the
percolation transition \cite{fluctuations}, which occurs (for this
realization) at $p=0.505$, close to the bulk classical percolation
critical point.

\begin{figure}[h!]
\centering
\includegraphics[width=8truecm]{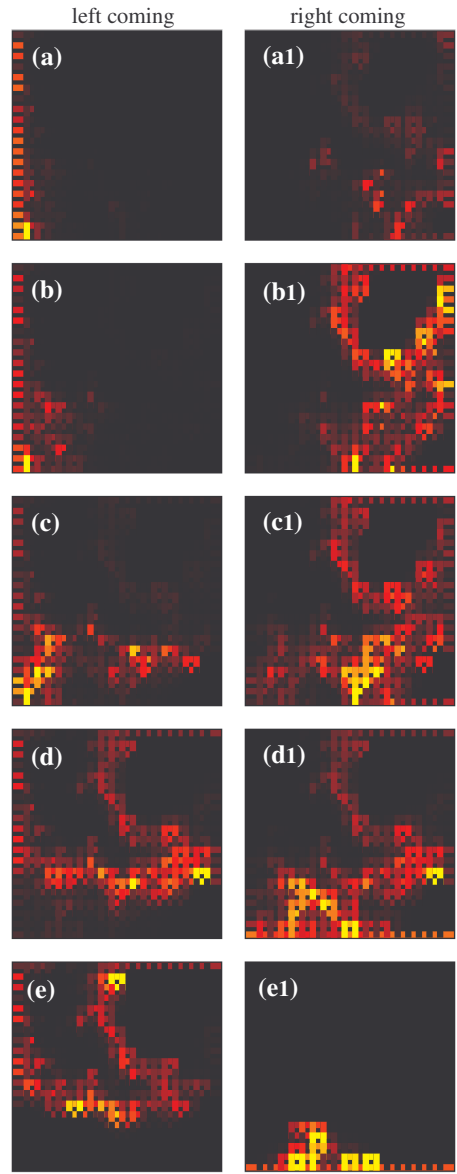}
\centering
\caption{\footnotesize Spatial current distribution when current is injected from the left (left column)
 or from the right (right column), for the concentrations depicted by arrows in Fig.~\ref{hotspots1}. In the
numeric calculation the current is injected from the upper-left-most or lower-right-most
link, corresponding to left- or right-injected current. As seen, while far from the transition
left- and right-coming currents are spatially separated, close to the transition the percolative nature of
the system causes a wide spatial current distribution. }
\label{currents} \end{figure}

The current distributions obtained for a specific realization of
disorder for different values of $p$  are depicted in
Fig.~\ref{currents} for two cases, when the current is injected
from the left (left column) and from the right (right column),
bright colors correspond to a high current. Each row corresponds a
concentration $p$ as denoted by arrows in Fig.~\ref{hotspots1}. As
can be seen, for low concentrations  (i.e. on the insulating side
of the transition, Fig.~\ref{hotspots1}(a)-(b)) the currents flow
in the system along some closed trajectory,
 returning back to the lead
they came from due to the low transmission.
In this realization the potential barrier separating left and right is located closer
to the left side of the sample, thus an electron injected from the left is reflected almost
immediately, while an electron injected from the right meanders through a larger part of
 the sample before being reflected.

Close to the transition the amount of current that passes through
the system from one side to the other is roughly equal to the
amount of back-scattered current, since the transmission of the
system is close to $T \approx 0.5$ (Fig.~\ref{hotspots1}(c)-(d)).
Here we find that the correlation between  left- and
right-originating current distributions is higher,
 and that the spatial distributions are broad, in accordance with
 percolation theory. Finally, for large concentrations
 (Fig.~\ref{hotspots1}(e)) the transmission of the system is
 perfect, and thus the current passes between the leads
 following some trajectory, which corresponds to a percolating
 path of perfect links. As in the case of low concentration,
 the chirality causes strong separation between the distribution
 of left- and right-coming currents.
A similar study, based on a tight-binding description,\cite{Cresti}
have resulted with similar spatial current
distributions, and has emphasized the role of the current chirality
in  the quantization of the Hall conductance and the vanishing
of the longitudinal resistance far from the transition. The left-right a-symmetry
in the presence of magnetic fields was also pointed out in Ref.~\onlinecite{metalidis}.

\subsection{Hot Spots} \label{subsec_hotspots}
In order to simulate the "hot spots" experiment,\cite{enslin} we
mimic the local gating  at a certain site by adding additional
energy to the potential barriers surrounding that site, up to a
distance of several lattice spacings. The conductance of the
original lattice is then compared with that of the perturbed lattice
for different values of the Fermi energy (corresponding to the
experimental change in the magnetic field, or filling factor).

\begin{figure}[h!]
\centering
\includegraphics[width=8truecm]{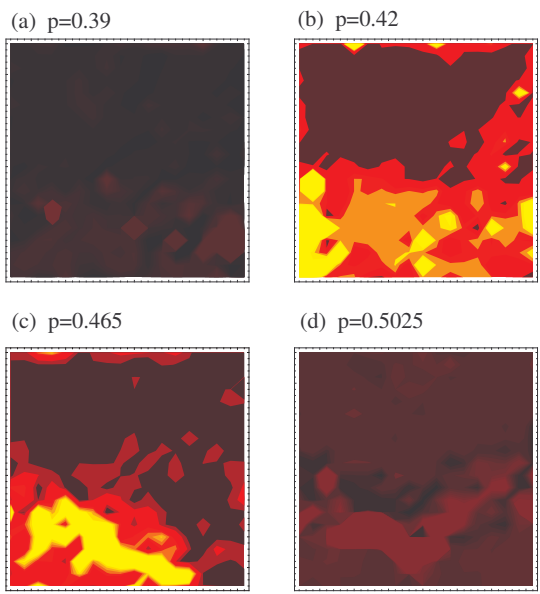}
\centering
\caption{\footnotesize Spatial mapping of the "hot spots"
in the conductance, for probabilities corresponding to the
arrows (b)-(e) in Fig.~\ref{hotspots1}. Far below the QH
transition (a) there are very few hot spots. Their number
and intensity increases
as one approaches the transition (b), is maximized at the
transition region (c) and again diminishes above the transition (d).}
\label{hotspots2} \end{figure}

In Fig.~\ref{hotspots2} the spatial distribution of the change in
the conductance is plotted for different values of $p$,
corresponding to the points denoted $b,c,d,e$ in
Fig.~\ref{hotspots1}. Brighter points correspond to the "hot
spots" of the current, for which there is a sizable change in the
conductance as that point in the lattice is gated by the AFM tip.
The numerical data are normalized to the largest conductance
change.

\begin{figure}[h!]
\centering
\includegraphics[width=7truecm]{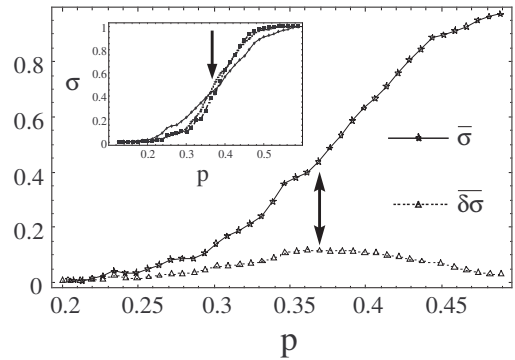}
\centering \caption{\footnotesize Conductance (stars) and absolute
value of the change in the conductance (triangles) as a function
of concentration $p$, averaged over all lattice sites and over 100
realizations. The arrow indicates the point of percolation,
defined as the point at which the disorder-average conductance is
independent of system length. The inset shows the conductance for
system lengths $L=10,15,20$. }
\label{hotspots3} \end{figure}

As deduced from Fig.~\ref{hotspots2}, both the amount and
intensity (i.e. how considerable the influence
 of depleting
sites from electrons on the conductance is) of the hot spots
increase as one approaches the QH transition, in agreement with the
experimental result. This may be seen more clearly by calculating
the absolute value of the
 conductance change $\del \si$, averaged
over all the lattice sites and over disorder. $\overline{\del \si}$ is
plotted in Fig.~\ref{hotspots3} for $100$ disorder realizations.
It is found that the maximal change in the conductance corresponds
to the percolation threshold, determined by the point at which the
conductance is length-independent, denoted by an arrow in the
inset of Fig.~\ref{hotspots3}.

\subsection{Effect of tip parameters}
Let us turn our attention to the role of the AFM tip.
Experimentally, the exact effect of the AFM tip on the sample is
unknown. However, it is
 reasonable to assume that the tip induces an increase in the potential energy in the area  underneath it,
 which affects electrons up
to a length $l_{eff}$ away from it.

In order to address this point theoretically, we note that the tip has two main tunable
 parameters, namely the potential difference (voltage)
between the tip and the sample, $E_\mathrm{tip}$, and the distance
between the tip and the sample. Changes in these experimental
parameters affect two different aspects of the model: (a) a change
in the offset of the energy in the links underneath the tip,
namely the local potential energy change induced by the tip, and
(b) a change in the effective length, $\leff$, over which
electrons feel the tip. While experimentally these two parameters
are both affected, to some degree, by the tip height and voltage,
theoretically one can study the change in each parameter
separately. To simulate these effects, we repeat the above
calculation, with a tip-induced exponential-enveloped change in
the local potential barrier on the links,
 $\D E=E_\mathrm{tip} e^{- d /l_\mathrm{eff}}$, where $d$ is the distance between the link and
 the position of the tip. In what follows we explore how changing either $\Etip$ or $\leff$ affects
the conductance change.

\begin{figure}[h!]
\centering
\includegraphics[width=8truecm]{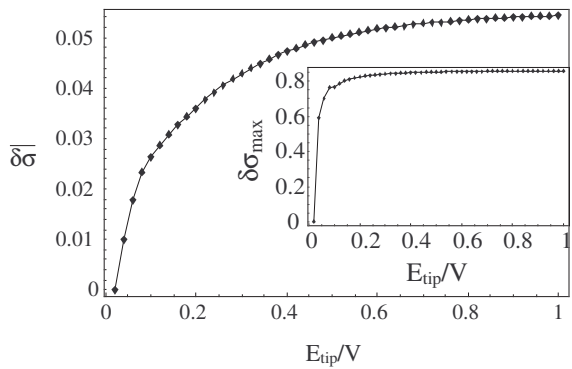}
\centering
\caption{\footnotesize Average change in conductance, $\overline{\del \si}$, as a function
of STM tip voltage $\Etip$, averaged over the entire sample. Inset : the same for the strongest
 "hot spot" in the sample.    }
\label{EtipChange} \end{figure}

Changing $\Etip$ does not have a significant effect on  the
spatial distribution of the hot spots, but only on their strength,
namely the conductance change induced by the tip. In
Fig.~\ref{EtipChange} we plot the average change in conductance
$\overline{\del \si}$ (averaged over the entire sample) as a
function of $\Etip$, varied from $\Etip=0$ to $\Etip=V$, that is
of the order of the band width. The calculation is performed for
concentration $p=0.42125$, where the number of hot spots is quite
large,  and $\leff=1$ (in units of lattice spacing). As seen,
$\overline{\del \si}$ increases monotonically with $\Etip$. In the
inset of Fig.~\ref{EtipChange} we plot the conductance change
$\del \sigma$ when the tip is over the strongest "hot spot" in the
sample, and again, a monotonic increase in $\del \si$ is observed.

Next we examine the effect of changing $\leff$. One may naively guess that increasing $\leff$ should result
in an increase in $\del \si$. However, due to quantum interference this may not always be the case, especially
far from the percolation transition, where the conductance change is rather small. In Fig.~\ref{LeffChange}
We plot the spatial map of hot-spots for $\Etip=V$ and $p=0.30875$ (which is close to the percolation
transition for this sample, the sample having a transmission $T=0.648$ for this concentration), for
 $\leff=1,2,..,8$ (in units of lattice constant). We find that although the number of hot-spots
 (indicated by bright colors) increases, their locations are changed, due to interference effects.

\begin{figure}[h!]
\centering
\includegraphics[width=8truecm]{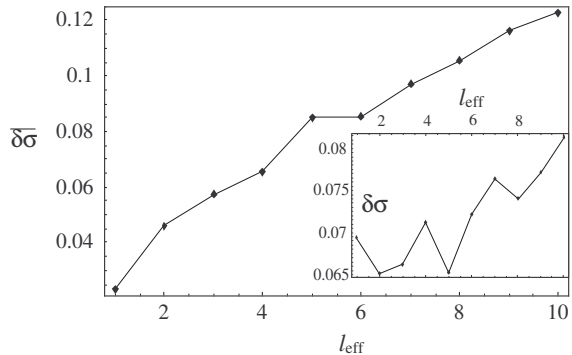}
\centering
\caption{\footnotesize Average conductance change, $\overline{\del \si}$, averaged
over the sample and over $100$ realizations of disorder as a function of $\leff$, showing a monotonic increase.
Inset : the local change $\del \si$ at a given position and realization of disorder as a function of $\leff$.
One sees that $\del \si$ may fluctuate due to quantum interference.  }
\label{LeffChange2} \end{figure}

To make this more qualitative, in Fig.~\ref{LeffChange2} we plot the average change in the conductance
 $\overline{\del \si}$, averaged
over the sample and over $100$ realizations of disorder, for concentration $p=0.5$ (i.e. with Fermi energy at
the center of the band), as a function of $\leff$. We find that indeed, on average the conductance change
 exhibits a monotonic increase. However, for a given realization and a given tip position,
changing $\leff$ (which corresponds to a change in the distance between tip and sample) may result
 in fluctuations in $\del \si$, as plotted in the inset of Fig.~\ref{LeffChange2}.
 \begin{widetext}
\begin{center}
\begin{figure}[h!]
\centering
\includegraphics[width=14truecm]{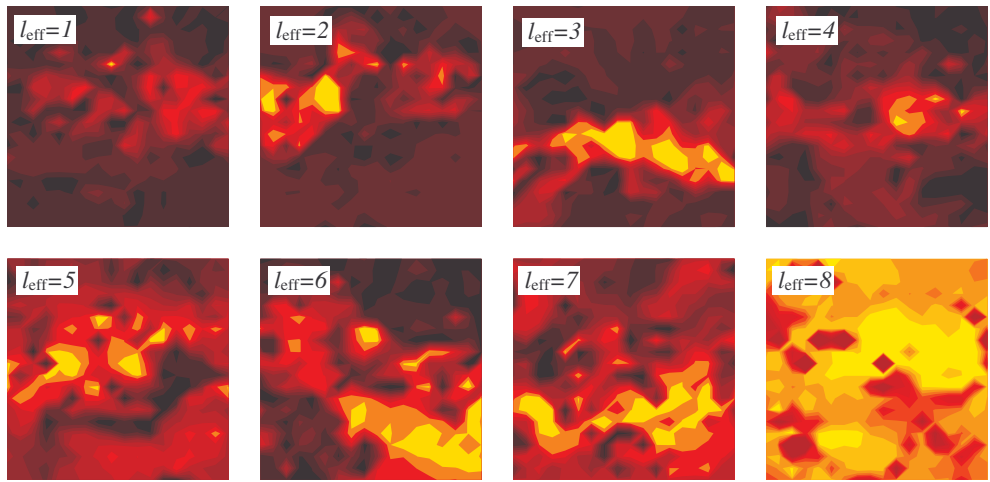}
\centering
\caption{\footnotesize Spatial map of hot-spots for different values of the STM tip effective length,
 $\leff=1,2,..,8$. }
\label{LeffChange} \end{figure}
\end{center}
\end{widetext}

\subsection{Relation Between Hot Spots and Current Distribution}

Next, we ask the question: are the "hot spots" observed in
experiment located at
 the extended, current carrying electronic states in the bulk ? As stated above, although one is inclined to
give a positive answer, the different symmetry of the hot-spots
and the current distribution points that they cannot be identical.
It is clear, however, that they are correlated, in a way we
discuss below.

\begin{figure}[h!]
\centering
\includegraphics[width=8truecm]{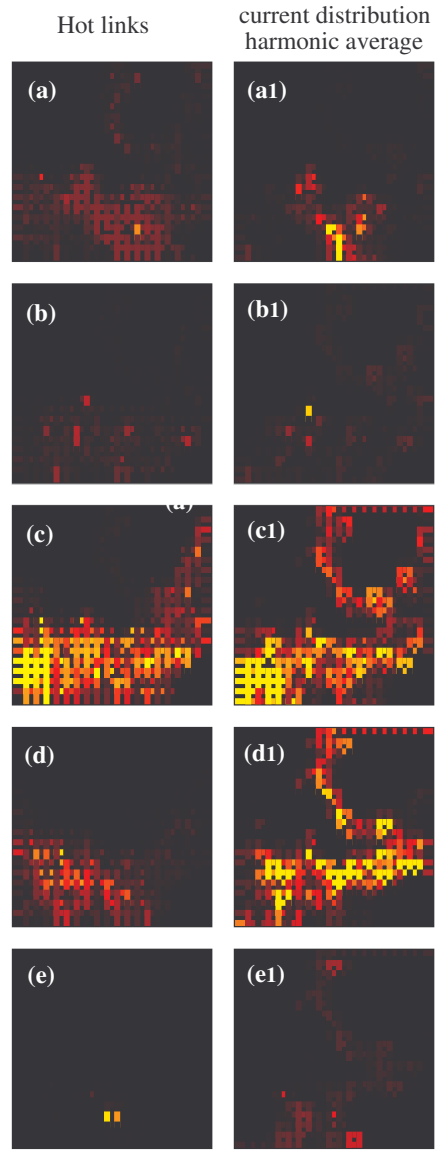}
\centering
\caption{\footnotesize Left column: a spatial image of the hot links,
 for the concentrations depicted by arrows in Fig.~\ref{hotspots1}, bright links correspond
to links with strong effect on the conductance. Right column: spatial image of the
harmonic mean of local currents obtained from two different
directions of voltage drop, for the same concentrations. Bright links
correspond to links in which current is considerable for both
directions of voltage drop. A clear
correlation between the left and right columns is visible. }
\label{HotLinks} \end{figure}

Let us examine the correlation between the hot-spots and current
distribution \cite{link}.  Since the strength and location of the
hot links is independent of the direction of the current
injection, it is clear that the hot links are {\sl not} located at
the points where most of the current passes, as these depend
sensitively on direction. Rather, it is found that the hot links are
located at points which hold appreciable currents for both
directions of current injection. To demonstrate this we plot on
the left column of Fig.~\ref{HotLinks} the spatial image of a
harmonic average of the local currents from the two directions of
current injection. Bright spots thus correspond to links in which
local current is significant in both directions of current injection.
On the right column we plot the spatial image of the "hot links",
that is, the conductance change due to setting the transmission on
each link to zero. Both columns are shown for the concentrations
corresponding to the arrows of Fig~.\ref{hotspots1}. One clearly
sees the correlation between the two images.

\begin{figure}[h!]
\centering
\includegraphics[width=7truecm]{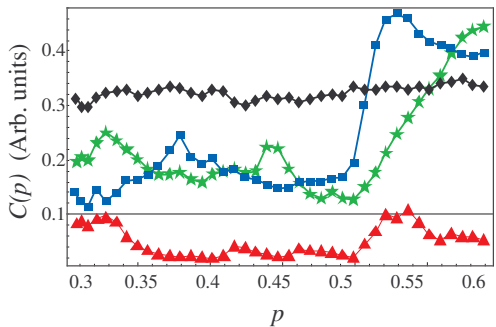}
\centering \caption{\footnotesize Correlation function $C(p)$ (see
text), as a function of concentration. The correlation is between
the hot-links and the harmonic-averaged current distribution
(triangles), the hot-links and the left-coming (stars) and
right-coming (squares) current distribution  and between the
hot-links and a random current distribution (diamonds). The lower
the value of $C(p)$, the higher the correlation. The low value of
the correlation function between the hot spots and the
harmonic-average, relative to the other correlations, demonstrate
that the hot spots do not directly reflect the total spatial
current distribution, but only the transport current, estimated by
the harmonic average.}
\label{corr} \end{figure}

In order to make this correlation more quantitative the
correlation function $C(p)$ is defined to be the square of difference between
the current and the conductance change, normalized and averaged
over the entire sample. That is, let $j_i$ be the normalized current in the $i$-th link
(either when the current is injected from the left, from the right or a average of the two),
and $\del \si_i$ be the normalized change in the conductance when the tip is placed over the $i$-th link (note
that both $j_i$ and $\del \si_i$ depend on the concentration $p$),
then $C(p)=\frac{1}{N} \sum_i |j_i -\del \si_i|^2$.
       The smaller $C(p)$ is, the higher is the
correlation between these distributions. In Fig.~\ref{corr} $C(p)$
 is plotted for different concentrations $p$, when correlating
between the hot links and the harmonic averaged current
distribution (triangles), the hot-links and the left-coming (stars)
and right-coming (squares) current distribution  and between the hot-links and a
random distribution (diamonds), that serves as a reference scale.
As seen, the hot-links are well correlated with the
harmonic-averaged current distribution, indicated by the low
values of $C(p)$. The correlation between the
hot-links and currents flowing from the left or from the right is much worse, and
actually resembles the correlation with a completely random
distribution for large concentrations.

\section{Summary}
In  this work a recent model for the QH
transition\cite{us} have been used to shed light on an experiment in which a
local AFM-tip induced local gating has been employed to study the
change in the resistance as a function of the tip location. The present
theoretical study demonstrates the existence of "hot spots",
regions with higher sensitivity to local gating. Our model,
consisting of a coherent mixture of perfect and quantum links,
qualitatively reproduces the experimental result.
 It was demonstrated that these "hot spots" do not lie
in the current-carrying paths, but rather on areas in the sample
where current flows both when it is injected from the left or from
the right, that is on the left-right symmetric parts of the
current carrying paths. Note that since the geometry employed in
this study corresponds to a two-terminal geometry, the
longitudinal conductance is trivially related to the Hall
conductance, and thus one expect a similar behavior of the "hot
spots" in the Hall conductance.

We conclude by noting that in order to verify our finding
experimentally, one should imply a non-destructive local probing
of the QH sample (that is, probing that does not affect the
conductance), e.g. local current-induced magnetic field sensing,
in addition to the above mentioned local AFM-tip gating, {\sl on
the same sample}. Such an experimental
 system may turn out to be useful
for measuring local currents in other systems, such as
disordered superconducting thin films, or Hall
samples in the fractional QH regime.

We thank T. Ihn for valuable discussions. This research is
partially supported by a grant from the Israeli Science Foundation
(ISF).


\begin{references}

\bibitem{QHEreview} For a recent review see
{\textit The Quantum Hall Effect},  D. Yoshioka (Springer, New York, 2002).


\bibitem{QPC}
B. J. Van Wees, H. van Houten, C. W. J. Beenakker, J. G. Williamson,
L. P. Kouwenhoven, D. van der Marel and C. T. Foxon, \prl {\bf 60}, 848 (1988);
D. A. Wharam, T. J. Thornton, R. Newbury, M. Pepper, H. Ahmed, J. E. F. Frost,
 D. G. Hasko, D. C. Peacock, D. A. Ritchie and G. A. C. Jones,
 J. Phys. C:Solid State Phys. {\bf 21}, L209 (1988).


\bibitem{edge}
B. I. Halperin, \prb {\bf 25}, 2185 (1982);
A. H. MacDonald, T. M. Rice and W. F. Brinkman, \prb {\bf 28}, 3648 (1983);
M. Buttiker, \prb {\bf 38}, 9375 (1988);
D. B. Chklovskii, B. I. Shklovskii and L. I. Glazman, \prb {\bf 46}, 4026 (1992).

\bibitem{bulk}
G. Diener and J. Collazo, J. Phys. C {\bf 21}, 305 (1988);
H. Hirai and S. Komiyama, \prb {\bf 49}, 14012 (1994);
K. Tsemekhman, V. Tsemekhman, C. Wexler and D. J. Thouless, Solid State Commun. {\bf 101}, 549 (1997).

\bibitem{palevski}
E. Yahel, A. Tsukernik,  A. Palevski,  and H. Shtrikman, \prl {\bf
81}, 5201 (1998)

\bibitem{imaging}
S. H. Tessmer, P. I. Glicofridis, R. C. Ashoori, L. S. Levitov, M. R. Melloch, Nature {\bf 392}, 51 (1998);
A. Yacoby, H.F. Hess, T.A. Fulton, L.N. Pfeiffer and K.W. West, Solid State
Communications {\bf 111}, 1 (1999); G. Finkelstein, P. I. Glicofridis, R. C. Ashoori, and
M. Shayegan, Science {\bf 289}, 90 (2000); K. L. McCormick, M. T. Woodside, M. Huang,
 M. Wu, P. L. McEuen, C. Duruoz and J. S. Harris, \prb {\bf 59}, 4654 (1999);
N. B. Zhitenev, T. A. Fulton, A. Yacoby, H. F. Hess, L. N. Pfeiffer and K. W. West, Nature
{\bf 404}, 473 (2000);


\bibitem{Topinka}
M. A. Topinka, B. J. LeRoy, S. E. J. Shaw, E. J. Heller, R. M. Westervelt,
 K. D. Maranowski, and A. C. Gossard, Science {\bf 29}, 289 (2000).

\bibitem{metalidis}
G. Metalidis and P. Bruno, \prb {\bf 72}, 235304 (2005).

\bibitem{enslin}
A. Kicin, A. Pioda, T. Ihn, K. Ensslin, D. C. Driscoll, and A. C.
Gossard \prb {\bf 70}, 205302 (2004).

\bibitem{fluctuations}
S. Cho and M. P. A. Fisher, \prb {\bf 55}, 1637 (1997.

\bibitem{percolation}
R. F. Kazarinov and S. Luryi, \prb {\bf 25}, 7626 (1982);
S. A. Trugman, \prb {\bf 27}, 7539 (1983).

\bibitem{huckenstein}
for a review see, e.g., B.Huckestein, \rmp{\bf 67}, 357(1995).


 \bibitem{QHEperc}
I.V.Kukushkin, V. I. Fal'ko, R. J. Haug, K. v. Klitzing and K. Eberl,
\prb{\bf 53},R13260 (1996);
A. A. Shashkin, V. T. Dolgopolov, G. V. Kravchenko,
M. Wendel, R. Schuster, J. P. Kotthaus, R. J. Haug, K. von Klitzing, K. Ploog,
H. Nickel and W. Schlapp, \prl {\bf 73},3141 (1994).

\bibitem{us}
Y. Dubi, Y. Meir and Y. Avishai, \prb{\bf 71},125311 (2005);
Y. Dubi, Y. Meir and Y. Avishai, \prl{\bf 94},156406 (2005).

\bibitem{square}
The lattice was taken to be square. Since the QH transition is universal and geometry-independent
we do not expect the lattice shape to affect the results.

\bibitem{Cresti}
A. Cresti, G. Grosso and G. P. Parravicini, \prb {\bf 69}, 233313 (2004).

\bibitem{link} In order to be able to compare links in the model,
 the calculation described in Sec.~\ref{subsec_hotspots} was repeated with the
minor change that instead of raising the energy of several
barriers around a certain lattice site,  we now calculate
the change in conductance due to cutting each link. This way
the "hot links" were mapped, which may
be compared to the distribution of currents.

\end{references}
\end{document}